\newcommand{\Msun}{~M_\odot}

\newcommand{\kms}{\rm ~km~s^{-1}}

 \documentclass[preprint]{aastex}

\shorttitle{INSTABILITIES AND CLUMPING IN TYPE Ia SUPERNOVA REMNANTS}
\shortauthors{WANG \& CHEVALIER}

\begin{document}

\title{INSTABILITIES AND CLUMPING IN TYPE Ia SUPERNOVA REMNANTS}
\author{Chih-Yueh Wang and Roger A. Chevalier}
\affil{Department of Astronomy, University of Virginia}
\affil{P.O. Box 3818, Charlottesville, VA 22903-0818}
\email{cw5b@virginia.edu, rac5x@virginia.edu}

\begin{abstract}

We present two-dimensional high-resolution hydrodynamical simulations 
in spherical polar coordinates of 
a Type Ia supernova  interacting 
with a constant density interstellar medium.
The ejecta are assumed to be freely expanding with
an exponential density profile.
The interaction gives rise to a double-shocked structure susceptible to 
 hydrodynamic instabilities. 
The  Rayleigh-Taylor instability initially grows,
but the Kelvin-Helmholtz instability 
takes over, producing  vortex rings.
Provided the simulation is initiated early in the evolution
with a  perturbation $\ga 1$\%,
the instability reaches its fully developed nonlinear strength
within 5 doubling times.
The further nonlinear evolution  does not depend on the initial conditions.
Considering the small initial radii of Type Ia supernovae, they
are likely to reach this fully developed phase.
The nonlinear instability initially evolves toward longer wavelengths
and eventually fades away when the reverse shock front is in the
flatter part of the supernova density distribution.
Based on  observations of  X-ray knots and the
protrusion in the southeast outline
of Tycho's supernova remnant,
we  include clumping in the ejecta because these features cannot
be explained by instabilities growing from small perturbations. 
The  clump interaction with the reverse shock 
induces Rayleigh-Taylor and Kelvin-Helmholtz instabilities on the clump surface
that facilitate fragmentation.  
In order to survive crushing and to have a bulging
effect on the forward shock, the clump's initial density ratio to
the surrounding ejecta must be
at least 100 for the conditions in Tycho's remnant.
The $^{56}$Ni bubble effect
may be  important for the development of clumpiness in the ejecta. 
The observed presence of an Fe clump would then require a
non-radioactive origin for this Fe, possibly $^{54}$Fe.
The large radial distance of the X-ray emitting Si and S ejecta from the
remnant center indicates that they were initially in clumps.

\end{abstract}

\keywords{hydrodynamics ---instability ---  
           supernova remnants 
          --- supernovae: general --- supernovae: individual (SN 1572)} 

\section{INTRODUCTION} \label{sec:intro}

Type Ia supernovae (SNe) are thought to arise from the thermonuclear
explosion of a  white dwarf that accretes mass from its binary
companion (Whelan \& Iben 1973; Chevalier 1981a; Nomoto, Thielemann,
\& Yokoi 1984; Harkness 1991). 
Both Chandrasekhar and sub-Chandrasekhar mass models have
been considered for the mass of the white dwarf at the time of
the explosion.
In the sub-Chandrasekhar mass models, He detonation occurs on
the surface of the white dwarf (Woosley \& Weaver 1994; Livne \& Arnett 1995).
Although these models have some attractive features, they
seem less able than Chandrasekhar mass models to explain the
light curves and spectra of Type Ia supernovae
(e.g., H\"oflich \& Khokhlov 1996).
In Chandrasekhar mass models, ignition is initiated in the
high density central regions of the white dwarf.
In one type of model, the burning front propagates subsonically
through the white dwarf.
A popular example of this deflagration model is the W7 model
of Nomoto et al. (1984),
which can approximately reproduce the observed light curves.
Branch et al. (1985) found that the model
 can reproduce the observed spectra if there is mixing of matter
with a velocity $> 8,000\kms$, although Harkness (1991), in more
detailed models, found that mixing may not be necessary.
In another class of Chandrasekhar mass 
models, the initial subsonic burning front,
or deflagration, makes a transition to a supersonic burning
front, or detonation.
This is the delayed detonation, or DD, model (Khokhlov 1991).

Although observations of extragalactic Type Ia supernovae have given
insights into the nature of the explosions,
a clear determination of the supernova model has not been possible.
Observations of Galactic supernova remnants provide another window on
the structure and composition of supernovae.
Likely remnants of Type Ia explosions are Tycho's supernova and
SN 1006.
Tycho's supernova (SN 1572)
appears to be a Type Ia supernova  
based on its reconstructed light curve (Baade 1945), although
it may have been subluminous   (van den Bergh 1993).
Schaefer (1996) noted that
the light curve allows a Type Ib origin,
but the lack of evidence for stellar wind or photoionization
effects in the surroundings of Tycho's supernova argues against
a massive star origin. 
At X-ray and radio wavelengths, Tycho shows a similar  morphology, 
characterized by a limb-brightened circular shell.
The outer edge is sharply defined and is interpreted 
as the forward shock front.
Seward, Gorenstein, \& Tucker (1983)  found evidence for 400 X-ray clumps
in the {\it Einstein} observatory image.
At optical wavelengths, an H$\alpha$ emission filament,
 which is thought to be from a fast shock 
moving into a surrounding  neutral material,
 is observed along the NE perimeter, supporting the shock interpretation. 
The remnant outline spans  $8'$ in diameter, 
corresponding to a linear radius of $\sim$ 3 pc for an
assumed distance of
2.5 kpc (Raymond 1984).
The HI absorption show that the environment surrounding Tycho is 
fairly uniform, without a prominent
density gradient (Reynoso et al. 1999).
This property makes it especially suitable for detailed study.

Hamilton et al. (1986) calculated a
one-dimensional hydrodynamical
model to study the
X-ray spectrum of Tycho's remnant and obtained good agreement with
spectra from the {\it Einstein} Observatory
and other
X-ray observatories.
They took a three-layer composition structure
and were successful in
reproducing the observed Si line emission.
However,  they chose a constant density profile for the supernova.
When this type of supernova density profile interacts with a constant
density ambient medium, the outer shocked layers (where Si was placed in the
model) end up dense and cool.
A constant density profile is probably a poor representation of
an exploded Type Ia supernova.
Dwarkadas \& Chevalier (1998, hereafter DC98) recently reviewed the types of
explosion models that reproduce the basic properties of Type Ia
supernovae and found that an exponential profile is the best
approximation for the density profile
overall, although there can be significant deviations
from this profile.
When such a profile interacts with a constant density medium, the
shocked supernova gas has an approximately constant temperature.
In order to reproduce the observed spectral properties of Tycho's
remnant, DC98 suggested that either the
supernova gas has dense clumps or there was an early phase of
interaction with a circumstellar medium.
This was on the basis of one-dimensional hydrodynamic models for the
interaction of a Type Ia supernova with a surrounding medium.

In order to make further progress, we extend the hydrodynamic modeling
effort to two dimensions. 
One motivation for this is the investigation of instabilities
that result from the deceleration of the supernova ejecta by the
ambient medium.
Chevalier, Blondin, \& Emmering (1992) previously  
investigated the interaction of
 a   power law supernova density profile  with
an ambient medium.
In this case, the flow approaches a self-similar state, although
turbulent motions are continuously fed by the instability.
In the case of an exponential profile, the flow is no longer self-similar;
the density profile effectively evolves from a  steep power law profile to
a flatter one.
Chevalier et al. (1992) investigated the instability for
a range of power law indices and found that the flow remained qualitatively
similar. 
In view of this, we expect the exponential case to be qualitatively
similar while the steep part of the density profile is
interacting with the surroundings, even though the flow is no longer 
self-similar.
Our aim is to determine the nature of the evolution.
Dwarkadas (2000) has already made some numerical studies of this
phenomenon.

Another part of our effort is to include density inhomogeneities 
in the freely expanding supernova
gas.
The {\it ASCA} observatory has made
possible the imaging of Tycho's remnant in moderately narrow X-ray
bands (Hwang \& Gotthelf 1997).
There are two X-ray emitting knots on the eastern side of
the remnant where there is a protrusion in the remnant's outline 
(Vancura et al. 1995; Hughes 1997; Hwang \& Gotthelf 1997).
One of these knots is strong in Si line emission and the other is strong
in Fe line emission.
With the {\it ROSAT} HRI, Hughes (1997) has measured the proper motion of
features in the X-ray emission.
He found that these two knots show approximately undecelerated motion.
On the other hand, radio observations over the whole remnant show
substantial deceleration, with $Vt/R\approx 0.47$ (Reynoso et al. 1997
and references therein), where $V$ is the shock velocity and $R$ is the radius.
The implication is that the knots represent not only an inhomogeneous
composition, but also higher density clumps that have not been
decelerated.
At a distance of 2.5 kpc, the velocities of the knots are about $8,300\kms$
(Hughes 1997).

The other well observed Type Ia supernova remnant, SN 1006, does not
show evidence for the expansion of X-ray emitting clumps.
However, absorption line studies of the Schweizer-Middleditch star behind
the remnant have provided a powerful probe of the supernova remnant
structure and composition (Wu et al. 1983, 1997; Fesen et al. 1988).
These observations show evidence for central cool Fe II that is
compatible with expectations for the unshocked Fe rich matter in
the central region of a Type Ia supernova, although the amount
of Fe is smaller than expected
(Hamilton et al. 1997 and references therein).
In addition, there is evidence for Si II absorption that is
redshifted over the velocity range $3,000-7,000\kms$ (Wu et al. 1997;
Hamilton et al 1997).
Fesen et al. (1988)  initially argued that the Si rich matter is in a clump
on the far side of SN 1006.
However, Hamilton et al. (1997) more recently argued for a spherically
symmetric model for the supernova ejecta in which there is a transition
from Fe to Si rich gas at a velocity of $5,600\kms$ and the supernova
remnant is considerably more extended on the back side because of
a low interstellar density in that direction.
They argue against a Si clump, but appear to assume that the clump
gas would have been shocked and cooled.
Another possibility is that low ionization absorption is due to
unshocked matter in the clump.
The clump may be moving at about $7,000\kms$ and matter stripped from
the clump may lead to the matter extending to lower velocities.
The remnant can then have a shape that is closer to spherical symmetry;
the presence of a clump implies that the ejecta are not in spherical
shells, but this is also indicated in Tycho's remnant.

Our aim here is to examine how inhomogeneities in the ejecta
interact with the decelerated interaction region.
The observed properties can place constraints on the nature
of the inhomogeneities, which can be important for the explosion models.

The plan of our paper is as follows.
In \S~2, we discuss the density structure of a Type Ia supernova
and consider mechanisms that might give rise to inhomogeneities
in the ejecta.
Our method is given in \S~3.
Results on the growth of instabilities from small perturbations
are in \S~4.
The evolution of nonlinear inhomogeneities is given
in \S~5.
Our results are discussed in the context of observations of
Tycho's supernova remnant in \S~6.
Implications for models for Type Ia
supernovae are in \S~7 and the conclusions are in \S~8.

\section{DENSITY STRUCTURE IN TYPE Ia SUPERNOVAE}

Soon after a supernova explosion, the ejecta are expected to be
freely expanding, with velocity $v=r/t$.
In this phase, DC98
found that an exponential density profile 
generally describes the density distribution 
obtained from numerical 1-D (one-dimensional) explosion models, including
 delayed detonation,  pulsating delayed
detonation,  He detonation, and  deflagration models. 
The exponential  density profile is given by
\begin{equation}
\rho_{SN} = A \exp(-v/v_e) \  t^{-3},
\end{equation}
where $A$ is a  constant 
and $v_e$ is the velocity scale height, which is determined
by the total ejecta mass $M$ and explosion energy $E$. 
This expression allows for spherical expansion;
an element of supernova gas moves with a constant
velocity and has its density drop as $t^{-3}$.
Integrating the density and energy density over  space and time
gives (DC98)
\begin{equation}
M = 4 \pi A v_e^3, \qquad E=48 \pi A v_e^5,
\end{equation}
or
\begin{equation}
v_e = (E/6M)^{1/2} = 2.44 \times 10^8 \ E_{51}^{1/2} \ 
\left({M\over M_{ch}}\right)^{-{1/2}} \ \  \rm{cm} \ \rm{s^{-1}},
\end{equation}
\begin{equation}
A = {6^{3/2}\over 8\pi}{M^{5/2}\over E^{3/2}} =
    7.67 \times 10^6 \ \left({M\over M_{ch}}\right)^{5/2} \ E_{51}^{- {3/2}}
\ \ \rm{g} \ \rm{s^3}\  \rm{cm^{-3}},
\end{equation}
where $M_{ch} \equiv 1.4 M_{\odot}$ is the Chandrasekhar mass and $E_{51}$ 
 is the explosion energy in units of $10^{51}$ ergs. 
A larger ratio of $E/M$ gives a larger velocity scale height and produces 
a flatter density profile at a given velocity. 

In some Type Ia models and core collapse models, the outer density
distribution has been described by a power law.
Then, the ejecta density   
is given by $\rho \propto v^{-n}t^{-3}$, where $v=r/t$ and $n$ is a constant $>5$.
If the ejecta interact with  a shallow power law medium $\rho \propto r^{-s}$
with $s<3$,  there exist self-similar solutions for the shocked
flow (Chevalier 1982a).
The problem contains only two  
independent dimensional parameters: the coefficients of the
supernova density and the external density. 
In the exponential case there is an additional parameter, 
the velocity scale height $v_e$, so that the problem is not
self-similar.

The exponential profile is an oversimplification of the density
profile, as can be seen in fig. 1 of DC98.
In a thermonuclear explosion, the elements of gas receive most
of their energy from the burning of that gas, as opposed to
a shock wave as in the case of core collapse supernovae.
The expansion of these heated elements is what leads to the
approximately exponential density distribution.
If there is incomplete burning of the material, the internal
energy per unit mass of the gas is lower that than in
neighboring regions where complete burning has taken place.
The gas is expected to evolve toward pressure equilibrium,
compressing the incompletely burned region.
The magnitude of the effect can be estimated by considering
the relative energy densities produced by different burning
processes, and allowing the gas with the greater burning to
adiabatically expand to come into pressure equilibrium with the
other gas.
The difference in burning a C/O mixture to Si or to Ni 
does not lead to a large density difference: $\la 40$\% higher
density in the Si gas.
We will find that this is not sufficient to explain the
clumping indicated in Tycho's remnant.
If some C or O is left unburned, a greater degree of compression
is expected, but this 
typically applies to only the outermost layers of the supernova.
This effect can apparently be seen in models (e.g., the W7 model density
distribution shown in Fig. 1 of Branch et al. 1985).

On a longer timescale, the density structure can be changed
by the Ni bubble effect.
The occurrence of SN 1987A brought the realization that the
energy release by radioactive $^{56}$Ni and $^{56}$Co after the initial
explosion could affect the density and composition structure within
the supernova (e.g., Woosley 1988).
Li, McCray, \& Sunyaev (1993) analyzed the Fe/Ni/Co lines
from SN 1987A and showed that the filling factor of Fe with
velocity $\la 2500\kms$ was $\ga 0.3$ although the mass
fraction of this element was only $\sim 0.01$ in the same region.
They attributed this effect to the expansion of gas heated
by radioactive power deposition.
Basko (1994)  made more detailed calculations of the expansion
effect.
Li et al. (1993) noted that the structure should show up in young
supernova remnants and briefly mentioned the Crab Nebula.
Chevalier (2000) contrasted the operation of the effect in
Type Ia and in core collapse supernovae, and we follow his discussion
of the Type Ia events here.

We examine the expansion of a spherical volume of radioactive
gas, following Basko (1994).
To get a crude estimate
of the effect, we assume that  immediately after the explosion,
the supernova gas is freely expanding and has constant density
throughout.
The velocity at the edge of a spherical clump of $^{56}$Ni
with mass $M_{\rm Ni}$ is
\begin{equation}
U_0=7.7\times 10^3 \left(M_{\rm Ni}\over 0.5\Msun\right)^{1/3}
\left(E\over 10^{51}{\rm~ergs}\right)^{1/2}
\left(M\over M_{ch}\right)^{-5/6}\kms.
\label{bubv}
\end{equation}
The deposition of radioactive energy in the $^{56}$Ni gas leads to 
expansion so that the velocity of the outer edge of the region
becomes $U_\infty$.
Conservation of energy leads to the equation
\begin{equation}
X^5-X^2={5Q\over U_0^2},
\label{radio}
\end{equation}
where $X=U_\infty/U_0$ and $Q$ is the energy per gram deposited
by radioactivity.
The assumption of full deposition of energy in the gas implies
$Q=3.69\times 10^{16}$ ergs g$^{-1}$ for $^{56}$Ni$\rightarrow ^{56}$Co
and $Q=7.87\times 10^{16}$ ergs g$^{-1}$ for $^{56}$Co$\rightarrow ^{56}$Fe.
The expansion is limited by the diffusion of either the $\gamma$-rays
or the photons out of the high pressure bubble so that the full
values of $Q$ listed above may not be achieved.
The bubble expansion is expected to drive a shell in the incompletely
burned matter on the outside.
Basko (1994) showed that the shell is subject to the Rayleigh-Taylor
instability, but that the instability does not have much
effect on the shell expansion.
The shell may break into clumps.
Equation (\ref{radio}) shows that if $U_0$ is large,
the expansion of the Ni bubble is relatively small and
the factor increase in the filling factor ($X^3$) is small.
Clumps can still be created in the swept up gas.
On the other hand, small regions of $^{56}$Ni have the potential
to substantially increase their filling factor in the supernova gas.

The reference values used in equation (\ref{bubv}) are those
thought to be typical of Type Ia supernovae.
These supernovae typically reach maximum light on a timescale of 15 days.
This is the timescale on which diffusion of the internal radiative
energy becomes effective, which has the effect of smoothing pressure
gradients and stopping Ni bubble expansion.
The mean life of $^{56}$Ni is 8.8 days and that of $^{56}$Co is 114 days, so
we estimate that the $^{56}$Ni radioactive
energy is deposited in the radioactive gas, but not that from $^{56}$Co.
After the expansion due to the radioactivity,
the outer velocity of the Fe-rich region is $8,400\kms$
and about $0.13\Msun$ of matter has been swept into a shell
around the Fe-rich region.
These numbers are approximate and will change for the density profile
of a specific supernova.

As discussed in \S~1, Tycho's remnant shows evidence for two
knots that have moved with undecelerated motion.
At a distance of 2.5 kpc, the velocities of the knots are about $8,300\kms$
(Hughes 1997).
This velocity is  consistent with that expected in a clumpy shell
formed by the expansion of radioactive gas.
However, Fe is expected in a clump only if it  formed as non-radioactive
Fe.
The Fe clump may have formed as $^{54}$Fe.  
X-ray spectra of Tycho's  remnant have been obtained with the result that,
in general, Fe line emitting gas is at a higher temperature and lower
density than Si line emitting gas (Hwang, Hughes, \& Petre 1998 and
references therein).
This is also expected as a result of the Ni bubble effect.

These considerations show that there is some evidence in Type Ia
supernova remnants for clumping surrounding a large Ni bubble.
The agreement between the knot velocities in Tycho's remnant and those
expected for the Ni bubble shell is probably fortuitous.
There is some evidence for both Tycho and
SN 1006 that they were subluminous supernovae (e.g., van den Bergh 1993;
Schaefer 1996), which would imply a smaller amount of $^{56}$Ni.
For the subluminous Type Ia SN 1991bg, Mazzali et al. (1997, and
references therein) estimate the synthesis of $0.07\Msun$ of $^{56}$Ni
extending to a velocity of $5,000\kms$.
This mass is consistent with the upper limit on the Fe mass ($< 0.16\Msun$)
found by Hamilton et al. (1997) for SN 1006.

The density contrast that is created by the nickel bubble is
not well determined.
In the calculations of Basko (1994), the density contrast at the
edge of the bubble is  a factor $\sim 5$ above the surroundings.
However, in the case of a Type Ia supernova with a relatively
much greater amount of radioactive Ni, clumps of non-radioactive
material may be immersed in the radioactive Ni and subjected to
a greater compression (H\"oflich 2000).
We have considered a simple model for this process in order to
estimate the compression.
We assume that the expanding supernova gas has a constant density
and that the free expansion during the first day is negligibly
affected by the addition of energy, as above.
The energy equation can be written as
\begin{equation}
{1\over \gamma -1}{dpV\over dt}= {dQ\over dt} -p{dV\over dt},
\label{energy}
\end{equation}
where $\gamma$ is the adiabatic index, 
$V$ is the volume, and $dQ/dt$ is the heat addition from radioactivity.
Over most of the evolution, the gas is radiation dominated and
$\gamma=4/3$, although the thermal pressure may be significant in
the initial phases.
Provided $t\la 1$ day, the power input from $^{56}$Ni is approximately
constant at $4.8\times 10^{10}$ ergs g$^{-1}$ s$^{-1}$.
Consideration of eq. (\ref{energy}) shows that in non-radioactive gas
the pressure falls as $t^{-4}$; this also initially applies to the
radioactive gas, but it  switches to $t^{-2}$ evolution (at $t\approx
2\times 10^3$s for typical parameters).
If the radioactive gas can keep non-radioactive gas at its pressure,
the compression of the non-radioactive gas increases as $t^{3/2}$
and can reach $\sim 200$ after a day.
However, a typical clump cannot be maintained in pressure equilibrium at
the high pressure and a compression wave is driven into the clump
from the outside.
When the radiative diffusion timescale becomes less than the age, there
is the possibility that the gas is more compressible and larger
density inhomogeneities are created.
This effect has been found in calculations of shock breakout from
a star in supernova models (e.g., Chevalier 1981b).
An important aspect of the effect is that radiation pressure dominates
the gas pressure, which is the case here.
These issues will have to be examined in detailed computations.

This discussion shows that it is possible that Type Ia supernovae
do contain inhomogeneities which could be observed in a supernova
remnant.
The Ni bubble effect occurs on a timescale of 10 days, or a scale
of $10^{15}$ cm for a velocity of $10^4\kms$.
The structure resulting from the Ni bubble should become frozen into
the ejecta at this early time.

\section{METHOD} \label{sec:basi}

We used the 2-D (two-dimensional)  code ZEUS2D 
based on a finite difference scheme to carry out the hydrodynamical 
simulations 
(Stone \& Norman 1992).
The code uses an artificial viscosity to smooth shock transitions.
We first ran 1-D simulations  
with an inner  exponential density profile (ejecta)
interacting with
an outer constant density ambient medium, starting  
2 weeks after the supernova explosion. 
The initial density distribution evolved into an intershock structure
consisting of  reverse-shocked ejecta in the inner region and a 
forward-shocked
ambient medium on the outside, separated by a contact discontinuity (DC98).
The 1-D intershock profile was then used to initiate 2-D simulations.
The inner ejecta gas was freely-expanding,
using an inflow boundary condition. 
The outer gas was at rest. 
Gas pressure in the unshocked gas was unimportant.
We neglected the effects of magnetism, heat conduction, and radiation.
Magnetic fields may play a role in suppressing
 heat conduction and viscosity, but
are not expected to be dynamically significant in the supernova remnant.
The radiative cooling time for the optically thin gas at an age
$\la 300$ yrs is  greater than the age of the remnant.
We assumed that the energy losses through X-ray emission are small
so that the dynamics of
the flow are not affected by gas cooling.
We used an adiabatic index 
$\gamma={5/ 3}$.

In 1-D runs, the starting dynamical age determines the
initial ejecta-ISM interface position. 
Simulations with different initial ages showed that 
the evolution quickly converges within a few radial doubling
times.
The solution is unique, independent of when the simulation starts
(see DC98).
More generally, since the ejecta density profile
only depends on its explosion mass and kinetic energy,
the interaction with
an ambient medium can be described by 
a set of scaling parameters $R'$, $V'$, $T'$
using $M$, $E$, and $\rho_{am}$ (see Truelove \& McKee 1999; DC98)
\begin{equation}
R' = \left( {3M \over 4 \pi \rho_{am}}\right)^{1/3}  \approx
2.19 \ \left({M\over M_{ch}}\right)^{1/3} \  n_0 ^{-1/3} \ \ {\rm pc},
\end{equation}
\begin{equation}
V' = \left({2E\over M}\right)^{1/2}  \approx 8.45 \times 10^3 \ \left({E_{51} \over
{M/M_{ch}}}\right)^{1/2}
\ \  {\rm km \  s^{-1}},
\end{equation}
\begin{equation}
T' = {R' \over V'} \approx 248 \ E_{51}^{-1/2} \ \left({M\over M_{ch}}\right)^{5/6} 
\ n_0^{-1/3} \ \ {\rm yr},
\end{equation}
where $n_0 = \rho_{am}/(2.34 \times 10^{-24}$ gm cm$^{-3}$)
and $\rho_{am}$ is the ambient density.
The dimensional variables $r$, $v$, $t$
can be expressed in terms of the nondimensional quantities:
$r'=r/R'$, $t'=t/T'$, and $v'=v/V'$.
Nondimensional solutions can be conveniently returned to dimensional ones  
by re-scaling, 
and new dimensional solutions for different 
$M$, $E$, and $\rho_{am}$ can  be calculated.
For a particular density distribution,
one evolutionary sequence in the nondimensional variables
 represents all possible dimensional solutions.

\begin{figure}
\epsscale{0.9}
\plotone{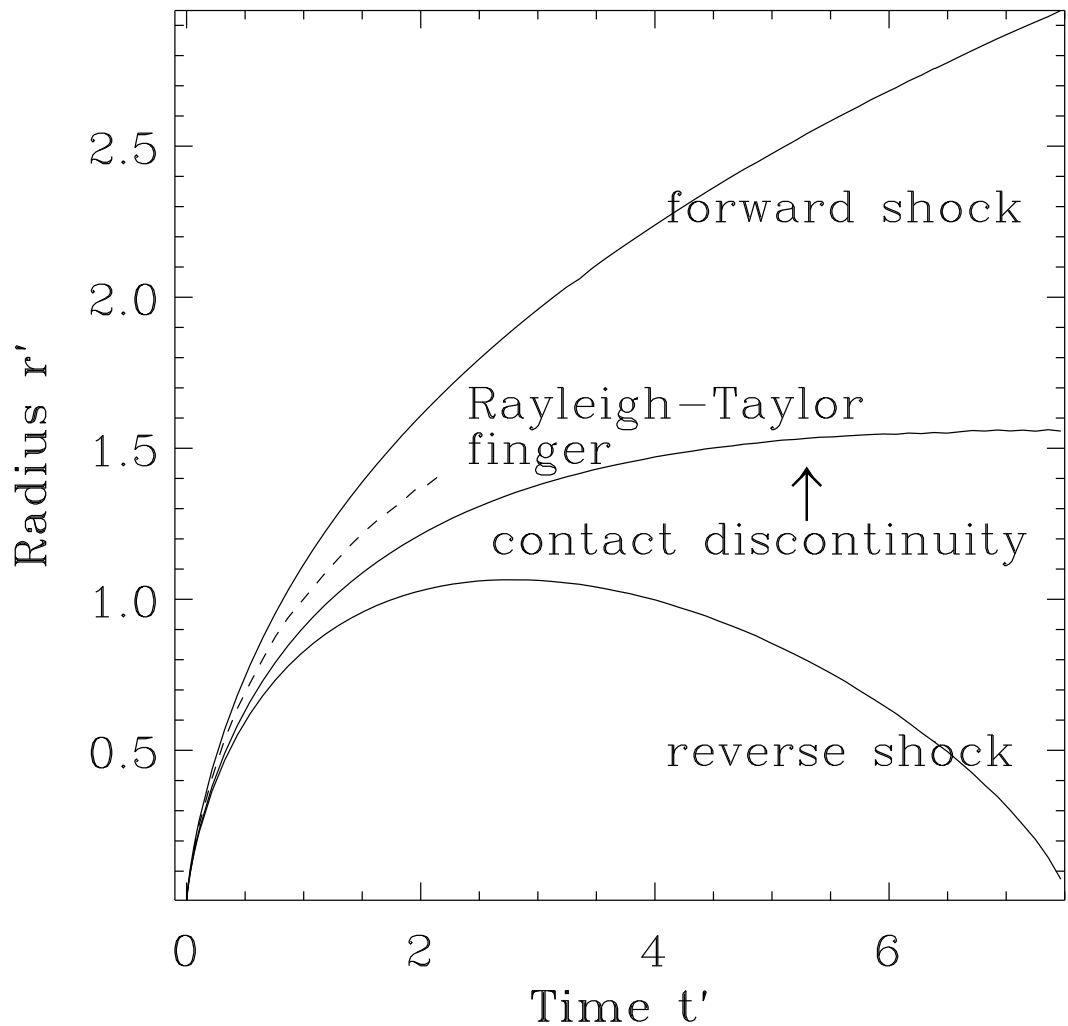}
\vspace{10mm}
\caption{One-dimensional evolution of the forward shock,
the contact discontinuity, and the reverse shock radius with time. The
farthest position to which the Rayleigh-Taylor fingers stretch
in two-dimensional simulations
is also plotted. The time and radius use normalized units given in the text.}
\label{fig.radius}
\end{figure}

For the exponential density profile, the 1-D solution (Fig. \ref{fig.radius})
entered the Sedov-Taylor self-similar blast wave evolution
with $ r \propto t^{0.4} $  
at $t' \gg 1$.
As the ejecta density decreased with time, the reverse shock, initially moving 
outward in
the stellar frame, began to move inward at $t' \approx 2.5 $.  
It reached the stellar 
center at $t' \approx 8$.


\section{INSTABILITIES FROM SMALL PERTURBATIONS} \label{sec:intro}

We simulated the Rayleigh-Taylor instabilities in the intershock region 
in two-dimensional spherical polar coordinates assuming $\phi$-symmetry.
The 2-D numerical grid was initialized with the 1-D solution at $t'=0.00054$
(0.134 yrs for the standard parameters $E_{51}=1$, $M=M_{ch}$, and
$n_0=1$) across the angular domain, with perturbation seeds in density,
pressure, and velocity placed between the contact discontinuity and the reverse
shock wave. The
grid was radially  expanding, following the intershock boundaries
 until the reverse shock radius turned
over to the stellar center. 
The evolution was tracked for five
orders of magnitude in time.
 Simulations initiated earlier  presented numerical problems
because the  higher  density contrast across the
contact discontinuity reduced the
numerical time step determined by the Courant condition.
 The 2-D grid was moved based on the 1-D results.
We used grid wiggling in the 
1-D simulations, tracking the forward and the reverse shocks.
In this process, the grid expands and contracts so that the shock locations
are kept at the same grid numbers.
We  fit the evolution of the radii and velocities 
with a linear function and a logarithmic polynomial,
respectively.
We then applied the 1-D smooth fits on the 2-D grid boundaries, 
avoiding significant numerical noise
coming from the wiggling scheme.   
The radial zone boundaries had a velocity distribution linearly varying with 
distance to the inner boundary, so that the zone spacing was kept uniform  
during the
expansion. 
We used 500 zones in the radial direction and varied the number of angular
zones; the radial zone spacing
was $3 \times 10^{-6}$ $R'$ at
$t'$=0.01,
and increased to $1.6\times 10^{-3}$ $R'$ at
$t'$=1.73.

We  investigated the growth of  instabilities 
with various grid resolutions and perturbations. 
To reduce the
computational expense,
only a fraction of a quadrant centered at $\theta=45^{\circ}$
was used. The run with the
finest resolution had 2000 angular zones in 1/2 of a quadrant.
The radial boundary conditions were inflow and outflow 
at the inner and outer sides, respectively. 
The angular boundaries were reflecting. 
We generally used a spherical harmonic function for the perturbation,
\begin{equation}
Y_{lm}(\theta,\phi) =  P_l^m(\cos\theta) e^{im\phi} ,
\end{equation}
where $m=0$ and the harmonic $l$ (perturbation mode) is even, considering the
$\phi$-symmetry and the reflection symmetry about the equator. 
The angular perturbation
is essentially  
 the associated Legendre polynomial $P_l^m$, which has
an increasing amplitude  toward
the polar axis. 
We  applied perturbations such that the amplitude was
 between 1\% and 50\% near 
$\theta=45^{\circ}$. The perturbed width was less than
40\% of the distance between the reverse shock and the contact discontinuity, 
about 2.5\% of the whole
intershock width.

\subsection {Evolution} 

Fig. (\ref{fig.evo}) illustrates the evolution of 
the instability perturbed by a 1\%, $l=100$
mode using 2000 angular zones on   1/2 of a quadrant centered
on $\theta=45^{\circ}$.
Small perturbations in the density grow 
and form spikes that protrude from the contact discontinuity
into the forward-shocked  region. As the remnant expands, 
the Rayleigh-Taylor instabilities 
continue to grow to form mushroom-shaped caps 
 while the stems become narrower. 
Secondary fingers  develop at the same time, so that there are more fingers
than the initial perturbation mode suggests. 
After reaching their maximum extent, the fingers fall to their sides and 
interfere with neighboring fingers.
 The forward growth of the caps is blocked by
the drag of the flow and the fingers do not extend farther from the
dense shell at the contact discontinuity.

In the later stages, 
the Kelvin-Helmholtz instability takes over at the mushroom caps,
creating vortex rings. 
The relative motion of flows between a finger and its surroundings bends
the stem and disrupts the flow. Since 
the effect of drag increases as the finger grows
outward, the continual shedding of mass at the finger's top 
eventually leaves
insuffient mass to overpower the drag of the countersteaming
flow. 
The mushroom cap falls off to the side, and the remaining filaments are
swept back. Vorticity  develops in the less dense regions 
left by the original mushroom caps.
Globally, the vortex rings gradually come to dominate over
the spikes, and only the stems remain recognizable.

The Rayleigh-Taylor instability consistently builds up fingers with long
wavelengths (low $l$). 
The dynamical stages of evolution  can thus be distinguished by the dominant
mode. We found four stages of evolution. The first stage 
is characterized by linear
growth of the Rayleigh-Taylor instability; 
the perturbations grow out from the initial seeds and
evolve towards a mushroom shape. Second, nonlinear growth of
secondary fingers among primary ones becomes important.
The fingers
multiply and become congested.
Third, 
mutual interference among fingers  increases the
wavelength of the dominant mode; vortex
rings develop as the Kelvin-Helmholtz instability becomes active, 
although new fingers still develop, as in a convective
roll. 
The instability is fully developed at this point.
In the final stage, the intershock density contrast is reduced
as the expansion decelerates; vortex rings and all
the other features in the flow are dispersed, and the instabilities fade.

\begin{figure}
\epsscale{0.9}
\plotone{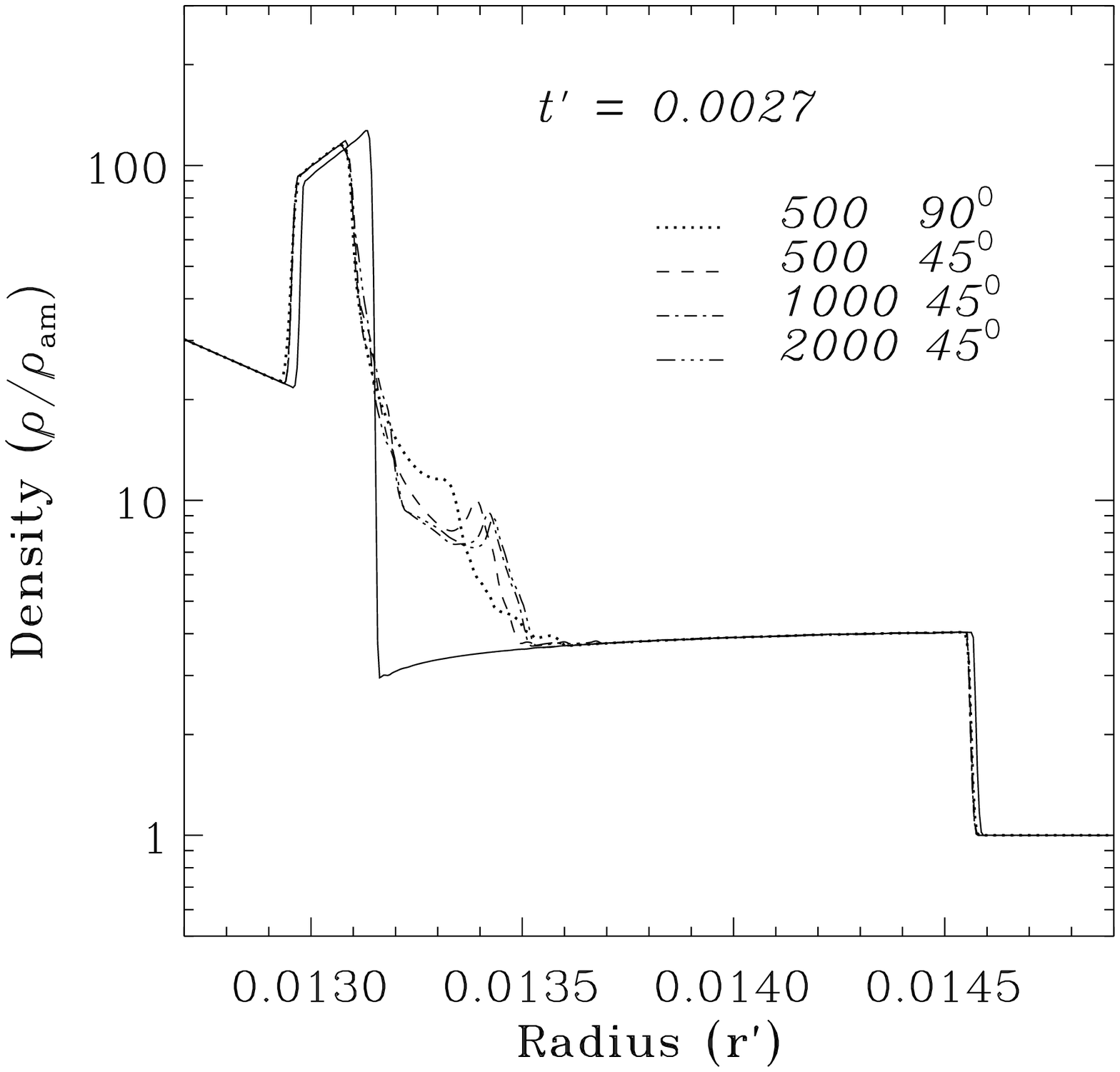}
\vspace{10mm}
\caption{Angle-averaged two-dimensional density distribution  
overplotted on the one-dimensional unperturbed solution in varying
grid resolutions with $l=100$ and 1\% perturbing amplitude.
The plot is at an early dynamical stage: $t'=0.0027$.}
\label{fig.angavg-early}
\end{figure}

\begin{figure}
\epsscale{0.9}
\plotone{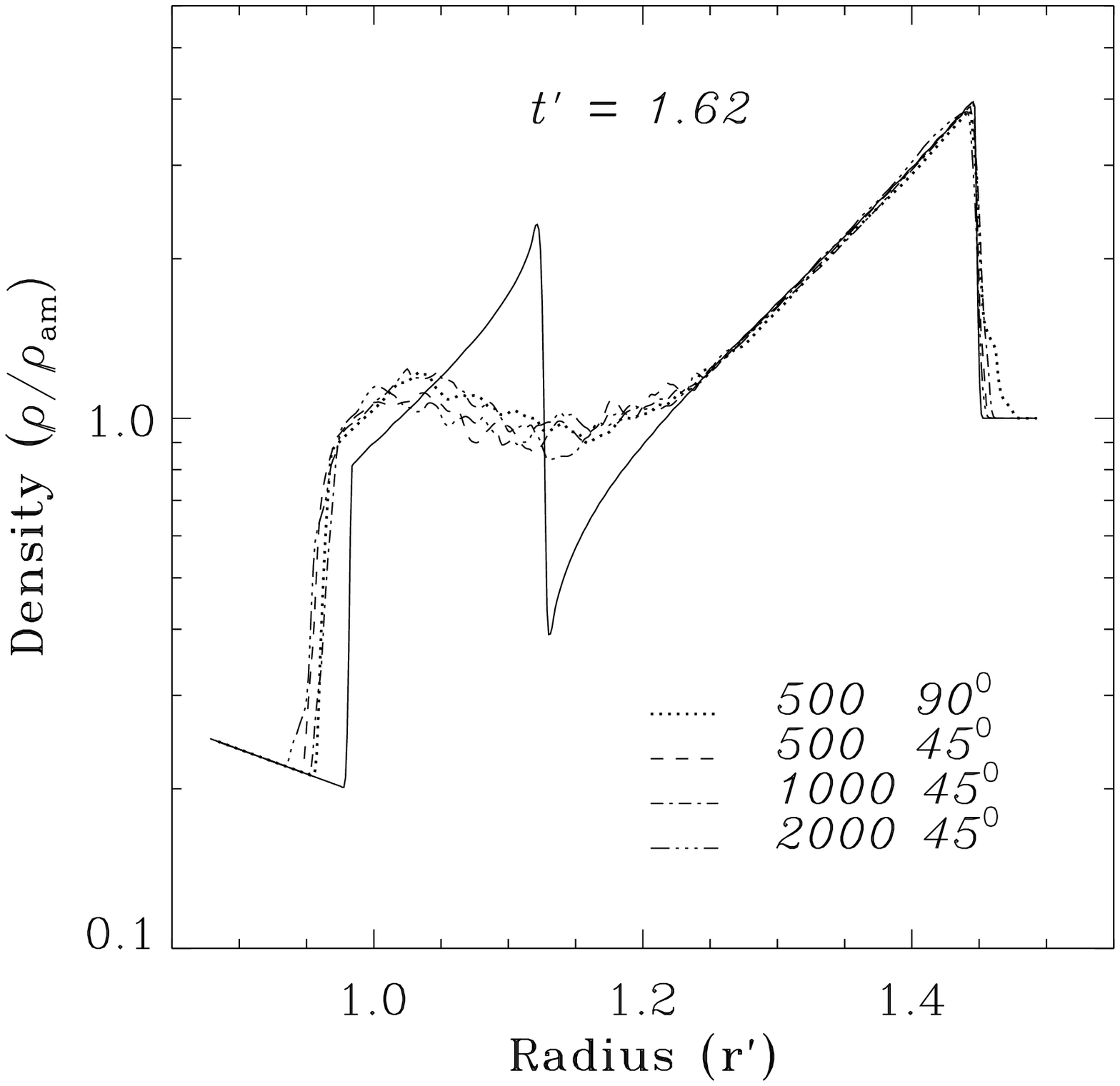}
\vspace{10mm}
\caption{Angle-averaged two-dimensional density distribution.  
The plot is at $t'=1.62$, corresponding to the present age of
Tycho's remnant.}
\label{fig.angavg-late}
\end{figure}

Figs. (\ref{fig.angavg-early}) and (\ref{fig.angavg-late}) show
 the angle-averaged density profiles with various resolutions
at an early phase and at $t'=1.6$ (approximately the present epoch
for Tycho). 
Compared to the 1-D
solution, the reverse shock front smears out somewhat
because it is corrugated, and
the density peak at the contact discontinuity is not present. 
The forward shock front is
less affected. 
Various
resolutions give basically the same profile, although
 finer grid resolutions reveal larger 
fluctuations. The late phase shows less density contrast and the decline 
of instabilities.

We compare the exponential case to the power
law case of $n=7$ with $s=0$ (interstellar medium) and $s=2$ (circumstellar wind)
in Chevalier et al. (1992).
In the power law case, the conditions are self-similar,
so that the evolution is 
quasi-steady.
Convection cells continually stir up a region
near  the contact discontinuity.
The density structure remains qualitatively similar with time. 
The difference in the two $n=7$ cases with different ambient media is that
for the $s=2$ case the density structure is characterized by 
groups of slender blobs with a pyramid outline,  
while for the $s=0$ case it is characterized by individual mushroom shapes.
The reason is that a constant density 
surrounding medium causes a smaller density
contrast across the the unstable region, and a
negative gradient of entropy. Thus, the $s=2$ case allows the initial blobs
 to grow and stretch out into
narrow fingers, contrary to the $s=0$ case in which the flow 
is blocked from propagating further and ends in a mushroom cap.

The $s=0$ power law case studied by Chevalier et al. (1992) shows
that a larger power law index $n$ increases the dominant mode $l$
and the evolution goes from quasi-steady convection to intermittent growth of
slender fingers, 
approaching in a sense the general profile of the low-$n$, $s=2$ cases. 
This is due to the larger density contrast 
between the shocked ambient medium and the shocked ejecta at larger $n$, 
and to the fact that
the gradients in the shocked ambient region become shallower.
The exponential case exhibits evolutionary properties
similar to the power law case with decreasing
index with time, in that longer wavelengths are built up throughout the
evolution after the initial growth stage.

\begin{figure}
\epsscale{0.9}
\plotone{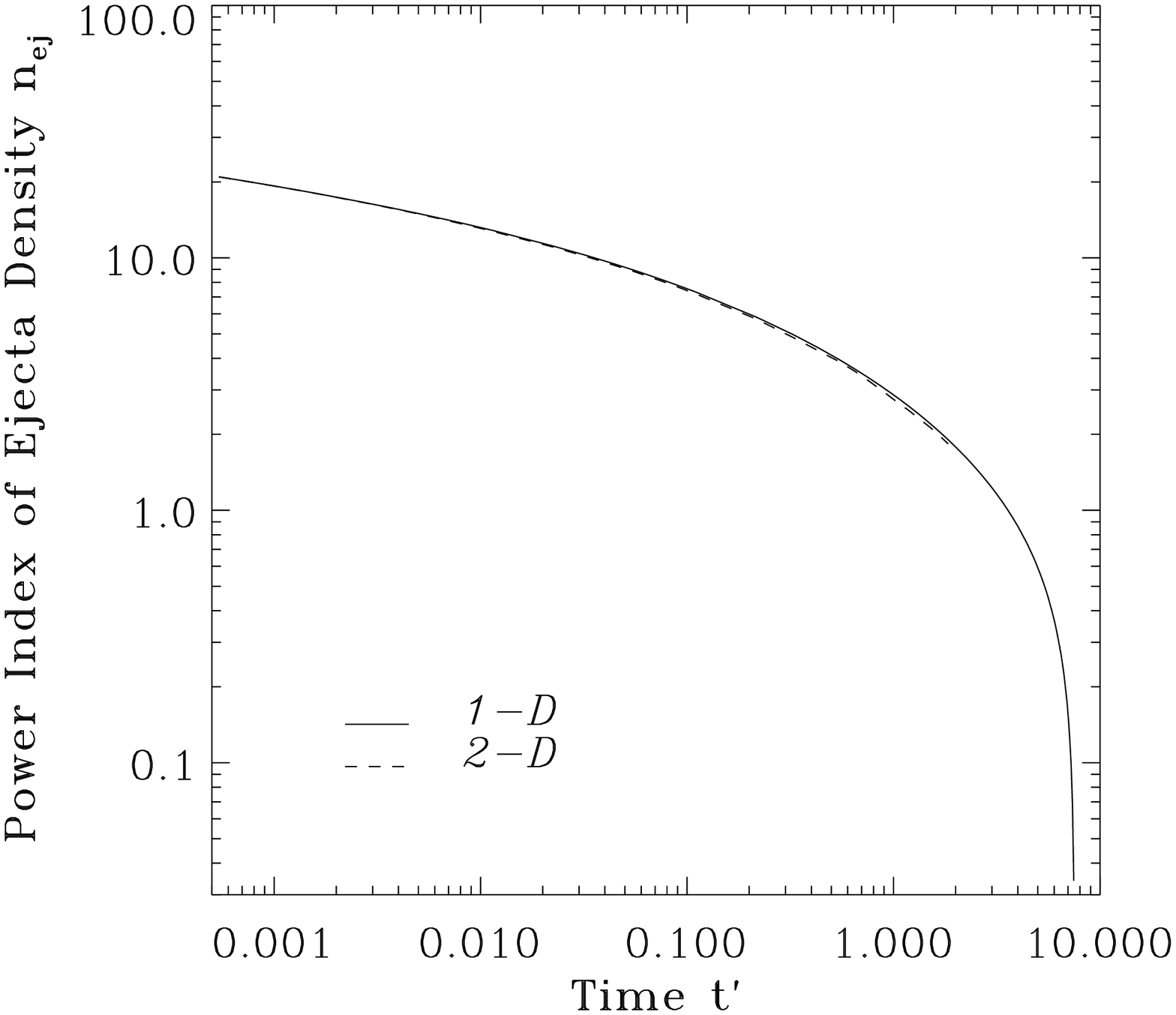}
\vspace{10mm}
\caption{Time evolution of the power law index of the ejecta density
profile immediately below
the reverse shock front from both the one-dimensional and two-dimensional runs.}
\label{fig.nej}
\end{figure}

We show in Fig. (\ref{fig.nej}) the approximate 
power law index $n_{ej} = - d\ln\rho / d\ln r$ in the unshocked ejecta
just inside the reverse shock 
for the exponential case.
In the 2-D simulations, the reverse shock surface is corrugated, but 
its position does not deviate significantly  from its position
in the 1-D case. 
The position of the contact discontinuity found by dimensional analysis
is $R_c \propto t^{(n-3)/(n-s)}$ for a power law ejecta density profile and
an ambient medium $\rho \propto r^{-s}$.
The power law index for $n = 5$ is the same as that for
a Sedov-Taylor blast wave.
For $n \le 5$ the forward shock is still expected to expand with
the Sedov law; however, the contact discontinuity expands more slowly
than $r \propto t^{0.4}$ and
 steadily distances itself from the forward
shock in the self-similar frame (Chevalier 1982a).
Self-similar solutions are no longer
possible and the reverse shock starts to move inward
towards the center. 
For the exponential case, we expect that the deceleration at the
contact discontinuity is reduced
when the reverse shock is in ejecta with
$d\ln\rho/d\ln r \le 5$. 
The instabilities should then lose the deceleration that drives them.
The index $n$ starts to fall below 5 at
$t' \ge 0.325$, while the turnover of the reverse shock actually occurs at
$t' \sim 2.5$; the motion of the reverse shock takes time
to respond to the changing density profile.

Both the density contrast and acceleration decline rapidly with time
in the exponential case;
the decline of the instabilities is an inevitable consequence.
At the end of our simulation ($t^{\prime}=2.16$), there is little continuing development
of the instability.

\subsection{Variation of the Perturbation  and the Resolution} \label{subsec:effpert}

The evolution of the instability is insensitive to the
initial linear perturbation.  
We have applied the initial perturbation 
in various areas: between the
contact discontinuity 
and the forward shock,
ahead the forward shock, and inside the reverse shock.
After the growth stage, the sizes of the unstable regions
are similar in all the cases.
Perturbations placed at the shocked ejecta near the contact 
discontinuity are the most effective in 
exciting the instability.  

On a low resolution grid of $500\times 200$ zones on  1/2 of a quadrant,
a 10\% perturbation with $l=100$ initially has
12 Rayleigh-Taylor fingers, which subsequently double in number.
The $l=50$ mode generates three secondary fingers between every two primary
fingers, resulting in 24 fingers like the $l=100$ mode (Fig. \ref{fig.resol}).
There appears to be
a preferred mode shortly after the linear growth,
independent of the initial conditions.
However, with finer resolution
more details between the primary fingers appear and
the presence of a dominant mode
becomes vague.
More highly resolved fingers appear to be more slender and sharp, and the overall
flow patterns thus appear different. 
On finer grids, the flow pattern approaches
the pyramid shape of an $n=7$, $s=2$ power law density
case; mushroom structures become
congested and twisted at the base near the contact discontinuity. 
Toward the
end of the evolution, a preferred mode in increasing wavelength is built
up independent of grid coarseness or the initial conditions. 
The presence of the
late-stage dominant mode can be understood as the exponential model having a
continually decreasing power law index, although limited resolution can also
fabricate a
longer  wavelength dominant mode. 
Our results are consistent with the studies of Rayleigh-Taylor
instability growth in a stellar explosion by  Fryxell, M\"uller and
Arnett (1991), who used a PPM (piecewise parabolic method)
code and cylindrical coordinates.

The morphology of the fingers on our higher resolution runs is similar to that
of Kane, Drake, \& Remington  (1999) using the PPM code PROMETHEUS. 
The results of Chevalier et al. (1992)
using the piecewise parabolic code VH-1, Jun \&
Norman (1996a) using ZEUS2D, and Fryxell
et al. (1991) using PROMETHEUS appear to have lower resolution than
our results and do not show the narrow structure that we find.
The high resolution calculation of M\"uller, Fryxell, \& Arnett (1991)
is closer to the morphology that we obtain.
Higher resolution calculations introduce shorter wavelength numerical noise
and allow small perturbations to develop.
Small scale structures have a faster
growth rate of the Rayleigh-Taylor instability in the linear regime,
and this can be seen in our simulations.
Fig. (\ref{fig.resol}) shows that the initial  growth of Rayleigh-Taylor fingers leads
to fingers with a larger extent at high resolution, presumably
because the smaller cross section gives less drag.
However, this trend may begin to reverse at the highest resolution because
the stems of the Rayleigh-Taylor fingers become more unstable
(see $t^{\prime}=0.003$ in Fig. \ref{fig.resol}).
As mentioned above, the extent of the unstable region shows little
dependence on resolution in the fully developed regime.
On finer grids,
the fingers in the nonlinear regime show an
increasing tendency to bend, which prompts stronger interaction among fingers
and produces more complex structures. 
As the resolution increases, the mushroom
caps break up into filaments;  this trend 
continues at our highest resolution. 
The clear
mushroom shapes found in lower resolution calculations
 may be an artifact of the limited resolution.
 
In independent work that was concurrent with our study,
Dwarkadas (2000) simulated the instabilities resulting from
the interaction of supernova ejecta (with an exponential profile)
with a surrounding medium.
He presented the evolution of the instability for a case in
which it grew from numerical noise.
This perturbation is much smaller than the cases that we considered
and the instability remained in the linear regime until the
late damping stage of the instability.
When he initiated the simulations with a 2 \% perturbation, his
results are consistent with ours.

We have also simulated the evolution for an initial perturbation
with 100\% amplitude.
After a transition phase during which the intershock region became
more than twice as broad as in the standard case, the evolution
returned to the standard case.
As found by Kane, Drake, \& Remington (1999) in simulations with
a power law supernova density profile, the standard evolutionary
track is robust.

\section{EVOLUTION OF CLUMPS}\label{sec:clump}

As discussed in \S~1, recent X-ray (Hughes 1997; 
Hwang \& Gotthelf 1997) and radio 
(Vel\'azquez et al. 1998) observations of Tycho's remnant show two knots 
juxtaposed
near the eastern edge.
The outer outline of radio (Vel\'azquez et al. 1998) and
X-ray (Vancura et al. 1995) emission, which presumably defines the forward
shock front, shows  an outward protrusion surrounding the knots.
The properties of undecelerated motion (Hughes 1997) and an outer
shock protrusion cannot be explained by instabilities generated
by linear pertubations.
The Rayleigh-Taylor fingers from linear perturbations are substantially
decelerated and do not affect the outer shock front.
Although our computations are two-dimensional, we believe that
going to three dimensions would not change this conclusion.
Rayleigh-Taylor fingers are known to show somewhat greater
growth in three dimensions during the initial growth phase
(e.g., Kane et al. 2000 and references therein).
However, in computations with power law supernova ejecta, Jun \& Norman (1996b)
find that the thickness of the mixing layer is very similar in two
and three dimensions when the instability is fully developed
(see their figs. 2 and 3).
The instability is probably limited by the properties of the
interaction region.
Jun \& Norman (1996b) do find that the stellar ejecta
show smaller scale structure in three dimensions.

We believe that the origin of the knots is best explained by 
nonlinear clumps of supernova ejecta 
expanding into the intershock region.
We thus simulated the  hydrodynamic evolution of  ejecta clumps
 interacting with the intershock gas. 
We represented clumps as denser spheres
superposed on the smooth exponential density profile. 
In the simulations, we initialized 
a clump at the polar angle $\theta = 45 ^\circ$
on  2-D computations like those described in the previous section.
In two dimensions, the sphere is in fact a 3-D torus around the polar axis. 
A 3-D sphere can be simulated by placing the clump
on the polar axis $\theta=0$ on a 2-D grid;
however, there are then singularity problems on the symmetry axis of the grid,
which lead to long Rayleigh-Taylor fingers on the axis due to the instabilities
discussed in the previous section.
Comparisons of the hydrodynamics of ISM clouds in spherical and toroidal
morphologies indicate that
 geometry contributes little to the interaction 
(Jun \& Jones 1999).
To examine the effect of geometry, we carried out clump interaction
simulations with clumps at $\theta=15^\circ$ and $\theta=80^\circ$;
the clump evolution did not substantially change.

Under the same boundary conditions as used for the instability computations, 
the clump was included in the freely expanding
ejecta. The clump first ran into the reverse shock
and then moved forward through the intershock region.
The initial clump as well as the ejecta were  cold compared to 
the high temperature shocked ejecta. 
The gas pressure was  unimportant,
 so that the size and the density contrast of the clump remained
unchanged until the clump expanded into the reverse shock front. 

The basic physics of the interaction is similar to that for the interaction
of nonradiative blast waves  with an interstellar cloud (Klein et al. 1994).
As the shock wave moves past the
 cloud, a reflected wave moves back into the shocked 
intercloud gas
and creates a bow shock. The incident shock also creates a transmitted 
`cloud shock' that moves into the cloud and crushes it.
The cloud shock propagates 
with a velocity $v_c \approx (\rho_i/\rho_c)^{1/2}v_s$,
where $\rho_i$, $\rho_c$, and $v_s$ are 
the density of the intercloud medium, the
density of the cloud, and the shock velocity in the intercloud medium,
respectively.
A larger density contrast between the cloud and the intercloud medium 
causes a smaller
velocity ratio of $v_c/v_s$, which helps the development of a shear flow,
and consequently the Kelvin-Helmholtz instability
at the cloud-intercloud interface.
The Richtmyer-Meshkov instability powered by the impulsive acceleration 
takes place due to the
impact of the shock wave on the upstream side of the 
cloud-intercloud surface. 
When the cloud shock exits the cloud, a rarefaction
wave moves back into the cloud and causes expansion. This acceleration
leads to the Rayleigh-Taylor instability on the upstream side of the cloud.
The combined
 instabilities lead to the  destruction of the cloud
on a timescale of several times of the cloud-crushing time, 
$t_{cc}=R_c/v_c$.

The clump-young supernova remnant interaction is more complicated than
the cloud-blast wave interaction
because of the structure of the intershock structure, 
including the pre-existing large scale instability.
The most important factor measuring the crushing
strength of a cloud-shock interaction
is the impact area per unit mass, which is inversely proportional to the initial 
density
contrast, $\chi = \rho_c/\rho_{i}$.  
We have explored the evolution of 
a single clump, varying its size, density contrast,
and initial impact time with the reverse shock. 
We examined four initial impact times, with the clump density contrast ranging
between 3 and 100,
and the clump size $a_0$ (as a fraction of the intershock width)
below 1/3 (Table~\ref{tabclump}).  
We estimate the  size of the Tycho X-ray knots 
to be about 1/5 the intershock width, although
this is uncertain because the clumps are poorly resolved and we see
only shocked parts of the clumps.
For a size below 30\% 
of the intershock width at the corresponding interaction time, we found that
({\it a}) an early clump initiated at $t'=0.011$  must have
a density contrast $\chi\ga 10$ to cause a protrusion
on the remnant outline; a clump with $\chi=30$  causes a protrusion but
the effect would have subsided long before the
present epoch for Tycho; 
({\it b}) a clump initiated at $t'=0.22$ with $\chi=50$ has a
bulging effect on the remnant outline, but the outline returns to spherical   
symmetry by $t'=1$; 
({\it c}) a later clump initiated at $t'=0.86$ with $\chi=100$ is not dense
enough to reach the forward shock at present epoch $t'=1.6$
for Tycho. A light clump with
$\chi=3$  develops vorticity when encountering the reverse shock
and is quickly destroyed.

Figs. (\ref{fig.G0mplt}) and (\ref{fig.k2mplt}) 
show the interaction with a single clump  for $\chi=100$ and two
different interaction times.
After passing through the reverse shock, the cloud  
gradually became flattened 
and curved like a crescent.
Material streamed out from the horns
of the crescent; the ram pressure difference between the axis
and the side of the cloud drove the mass loss.
The  pressure near the cloud axis was higher because of the additional
ram pressure on the front face of the clump. At the rear of the clump
the flow became turbulent and
left a trail of vorticies. 
The clump expanded laterally
as it approached the forward shock. 
The front of the clump  snowplowed the material ahead.
The  forward shock wave was distorted and 
a bulge  formed.
The material on the shock front 
became distributed into two lumps that gradually receded  
from each other.
As the  expansion of the supernova remnant continued, the bulge
as well
as the crescent clump immediately behind it
lost their identities. 
Even a clump with $\chi=100$ could not penetrate 
the forward shock when initiated at a later time (Fig. \ref{fig.k2mplt}), 
but snowplowed the material ahead and caused a protrusion.

The  velocity of a shocked clump is determined by the drag of the surrounding
material. A shocked clump  becomes comoving with the
postshock flow in approximately the
 time that the cloud sweeps up a column density
of intercloud gas equal to its initial column density.
The drag time is proportional to  $t_{drag} \sim \chi^{1/2} t_{cc}$,
so that a denser clump travels faster  
in the remnant.
The lateral expansion of a shocked clump into a  cresent shape increases the 
cross section area and the drag, significantly decelerating the clump.

For an exponential density profile of the ejecta, the clump
evolution depends on the time of clump interaction because of
the evolution of the supernova remnant.
An earlier clump requires less density contrast to have a protrusion effect on
the remnant outline (Fig. \ref{fig.G0mplt}).  
Early interactions give the clump
a crescent shape without
the action of the Rayleigh-Taylor instability on the clump's front 
because the higher
flow velocity in the remnant delays
the rarefaction wave traveling back upstream as
the cloud-shock exits the clump.  
Later clumps tend to
develop instabilities on the clump surface.
The exponential model gives a larger velocity difference 
and a larger density difference between the ejecta and
the reverse shock front at earlier epochs. 
The early clumps rapidly move into a lower density medium and
are thus more robust, for a given value of $\chi$.

For $n_o=0.8$ cm$^{-3}$, $E=10^{51}$ ergs, and $M=1.4\Msun$
(DC98), Tycho's remnant has
$t'=1.6$ at its present age of 427 years.
If this model applies and the knots are in undecelerated motion,
they must have crossed the reverse shock front at $t'\approx 1.2$.
This places Tycho in the regime where $\chi\ga 100$ is required.

In addition to the isolated knots, we considered groups of clumps.
The Ni bubble process is likely to create a shell of clumpy ejecta and,
as discussed in \S~6, the observations of ejecta in Tycho's remnant
suggest that widespread clumpiness may be present.
In Fig. (\ref{fig.band}) we show two simulations with multiple clumps.
In the top one, there is one band of clumps and in the second, there are
two bands.
The evolution of each forward clump is similar to that in the isolated clump
case, but they combine to push the forward shock wave to larger radius.
The ejecta move out to a relatively large radius, as may be required
in Tycho's remnant.

\section{THE REMNANT OF TYCHO'S SUPERNOVA} \label{sec:discussion}

As discussed in \S~1, Tycho's supernova was likely to be of Type Ia.
The expansion rate of the remnant has been measured in   the
radio, giving an expansion parameter
$m=0.47\pm 0.05$ (Strom, Shaver, \& Goss 1982),
$0.462\pm 0.024$ (Tan \& Gull 1985),
and $0.471\pm0.028$ (Reynoso et al. 1997).
The optical filament yields an  expansion parameter of
 $ 0.39\pm0.01$
(Kamper \& van den Bergh 1978), close to the adiabatic blast wave case. 
The slight discrepancy in these measurements can be 
explained by sampling variance; 
the optical measurement samples only the densest regions where
the emissivity is the highest.
HI observations do indicate a higher surrounding density to the
NE (Reynoso et al. 1999).
These radio expansion
values indicate a global pre-Sedov stage and exclude a 
surrounding circumstellar medium, which would give an expansion parameter
$\sim 0.7$ (Chevalier 1982b; DC98).
The exponential model is  successful in reproducing the observed
shock position and deceleration of Tycho's remnant.
At Tycho's present age of 427 years ($t'=1.7$)   
the model places the reverse shock and
the forward shock at  radii of 2.08 pc ($r'=0.98$) and 3.09 pc ($r'=1.45$),
while they decelerate with  expansion parameters of 
$m=0.15$ and $m=0.47$, 
respectively.

The radio observations of Tycho by VLA  by Reynoso et al. (1997) 
 show that the NE and particularly the adjacent SE parts protrude
from the circular outline.   The SE protrusion corresponds 
to the two X-ray knots.
Near the sharp NE  edge,
Vel\'azquez et al. (1998) note the presence of regularly spaced structure.
The spatial regularity  
is interpreted  by Vel\'azquez et al. as
 Rayleigh-Taylor fingers  still in the linear regime
with a mode $l \sim$ 30;  they attribute the preferred wavenumber 
 to the effects of  viscosity. 
However, the importance of viscosity is uncertain because magnetic fields
can inhibit its effect.
We believe that a linear regime for the 
Rayleigh-Taylor instability is unlikely.  
Regardless of the supernova density model,  
the shock front has expanded 
by a sufficently large factor to allow the
instabilities to evolve to their saturated phase if there are
 initial perturbations with amplitude $\ga 1$\%.

The regular structure observed in the radio image
by Vel\'aquez et al. (1998) is suggestive
of a Rayleigh-Taylor instability, but there are problems 
with this interpretation in the
context of our models.
One is that at $t'=1.6$, the region occupied by the unstable
ejecta does not extend beyond 85 \% of the remnant radius.
The structure observed by Vel\'aquez et al. is typically at
a larger radius and in fact extends to the edge of the remnant.
Another problem is that the instability at this late stage
does not have a clear regular finger structure in our models, but is dominated
by vortex rings with an irregular distribution.
The large radius of the structure suggests that it may
be connected to perturbations in the surrounding medium as opposed
to the ejecta.
However, the outer outline of the shock front is smooth on the
scale of the structure, as imaged at both optical and radio wavelengths.
In addition, the X-ray image of Tycho's remnant (e.g., Vancura et al. 1995),
which is dominated by line emission, shows strong emission at a
large fractional radius.
Because the image is likely to be dominated by emission from the ejecta
(Vancura et al. 1995; Hwang \& Gotthelf 1997),
the evidence is that the outer structure is related to the ejecta.

We believe that a solution to this problem is indicated by the presence
of the X-ray knots with undecelerated motion.
Two knots are clearly observed in one section, but it is likely that
they represent a more widespread phenomenon.
We suggest that there was a  spherical shell of clumpy ejecta at a velocity
$\sim 6,000-8,000\kms$ in the freely expanding ejecta
that gives rise to the clumpy structure.
Kane et al. (1999) previously suggested that the radio structure
observed in Tycho near the forward shock wave
may be related to nonlinear density variations
in the freely expanding ejecta.
There are reasons to believe that the clumps are restricted to a region
of the ejecta and not spread through all velocities.
First, the deceleration parameter measured at X-ray wavelengths is
significantly larger than that measured at radio or optical
wavelengths (Hughes 1997).
If the clumps are widespread, they should come and go in a band that
expands as does the emission at lower wavelengths.
However, a band of clumps could give the observed difference
in deceleration parameters.
Another reason is that in SN 1006, the Fe II absorption line profiles
are consistent with a distribution of gas at velocities $\la 5,000\kms$
like that in the W7 model (Fesen et al.
 1988), which has an approximately exponential
density distribution.
The Si line shows absorption at redshifted velocities $3,000-7,000\kms$,
which can be attributed to moderately high velocity Si interacting
with the surrounding medium.
There is no blueshifted Si absorption.
The data are consistent with the picture of smoothly distributed Fe on
the inside, outside of which is Si gas that may be clumpy.

The presence of a band of clumps is expected in the Ni bubble
scenario for clump formation (see \S~2).
One prediction of this model is that the outer ejecta emission
should be very clumpy when observed at high X-ray resolution
(see Fig. (\ref{fig.band}).
In addition to this component, there should be emission from
a smaller radius, near the reverse shock front.
This emission should be representative of lower density, higher
temperature gas.
We believe that the emission in the Fe K line, which is at
a smaller radius (Hwang \& Gotthelf 1997), is  primarily from gas close to
the reverse shock wave.
This gas should have some structure because of the earlier
instabilities, but much less than the outer clumpy ejecta.

Our computations showed that in order for knots to survive to close
to the outer shock wave, they must have an initial density contrast
with respect to their surroundings of $\sim 100$.
Hwang et al. (1998) in fact find that the ionization time ($n_e t$,
where $t$ is the time since the gas was heated) is 100 times larger
for Si than for Fe, based on line emission from the entire remnant.
The time since heating should be comparable for the two elements,
so this result is consistent with the Si being primarily in shocked clumps
(created outside the Ni bubble) and the Fe primarily in interclump
gas (inside the Ni bubble).
However, the situation is complicated by the fact that there is evidence
for some Fe in knots, which we argue had a nonradiogenic origin
(e.g., $^{54}$Fe).
In addition, we estimated in \S~2 that $\sim 0.13\Msun$ of Si could
be swept into a shell around the Ni bubble, so there could be
some Si emission from the ejecta that is not clumped.
Investigation of these issues will require spatially resolved
X-ray spectroscopy, as will be possible with the {\it Chandra}
and {\it XMM} observatories.

\section{IMPLICATIONS FOR TYPE Ia SUPERNOVA MODELS}

One result from observations of the Type Ia remnants is a lower
limit on the velocity of Si in the explosion.
For Tycho's supernova, there is an undecelerated knot of Si moving
at $8,300 (D/2.5{\rm~kpc})\kms$, where $D$ is the distance.
In SN 1006, there is Si freely expanding with a minimum velocity
in the range $5,600-7,000\kms$.
In both of these cases, the velocities appear to be less than
the minimum velocity of the Si-dominant layer in the W7
deflagration model, about $10,000 \kms$ (Thielemann et al. 1986).
However, if these are the remnants of subluminous supernovae,
the W7 model may not be the appropriate one.
Also, the initial burning front is unlikely to be precisely
spherically symmetric, so that some material that has not
burned to $^{56}$Ni may lag behind.

Our most significant results involve the evidence for clumping
in Type Ia supernovae.  
We have found that density contrasts $\ga 100$ are necessary to
reproduce the large radius of X-ray emitting clumps and their velocities.
A plausible mechanism for the formation of the clumps is the
action of an expanding Ni bubble, although calculations of
this process have not yet shown that such large compressions
are feasible.
We estimate that $\sim 0.1\Msun$ of matter can be swept into
clumps around the Ni bubble.
For the two undecelerated X-ray emitting clumps in Tycho's remnant,
Hughes (1997) estimates a mass of order $0.002\Msun$ for the
Si $+$ S clump and $0.0004\Msun$ for the
Fe clump, so there can be many comparable clumps.
The presence of Fe in a clump requires that it have
a non-radioactive origin.
The 1-dimensional calculations of nucleosynthesis associated
with the W7 Type Ia model (Thielemann et al. 1986; Iwamoto et al. 1999)
show that $^{54}$Fe is present along with Si and S in the region
immediately outside of the $^{56}$Ni region.
However, the $^{54}$Fe does not dominate the composition, unlike what is
indicated by the observations of the Fe clump in Tycho.
The central region of the W7 model is composed of stable $^{56}$Fe and
$^{54}$Fe, but this gas is at too low a velocity to be compatible
with the Tycho clump.
Thus, the synthesis of fast material dominated by non-radioactive
Fe remains to be demonstrated.

\section{CONCLUSIONS}\label{sec:conclu}

We used 2-D hydrodynamical simulations in spherical polar coordinates
to investigate   
instabilities and clumpiness in a Type Ia supernova remnant.  We adopted an
exponential density  profile for the smooth supernova ejecta and
a constant density for the surrounding (interstellar) medium.
The exponential
density profile gives the best general approximation for the supernova ejecta
in detailed explosion models, as opposed to a power law density profile 
for which  self-similar solutions are available.
The  region between the forward shock in the interstellar medium
and the reverse shock in the ejecta is  Rayleigh-Taylor unstable.
An initial perturbation grows to nonlinear strength, develops
an increasing characteristic wavelength, and fades as the remnant
evolves to the Sedov blast wave regime.
The characteristic Rayleigh-Taylor mushroom caps are replaced by vortex rings from
the Kelvin-Helmholtz instability, and
turbulence appearing in vorticies remains dominant throughout the rest of the 
evolution.
In these late phases, the depletion of 
the kinetic energy in the inner ejecta as 
the ejecta density profile continually weakens the acceleration
needed to power the Rayleigh-Taylor instability. 
A sequence of simulations with increasing grid resolution shows the 
Rayleigh-Taylor fingers become increasingly
slender and sharp and tend to bend,  limited only by the grid resolution.  

The structure that is formed as a result of the Rayleigh-Taylor
instability is decelerated and is confined to a region $\la 85$\%
of the remnant radius for a remnant with the dynamical age of
Tycho's remnant.
These properties are not consistent with the knots observed in
Tycho's remnant.
We thus considered a clump in the diffuse supernova ejecta   
with the aim of reproducing the properties
of the the SE X-ray knots and the related protrusion on
the forward shock outline.
In the clump-shock interaction,
the density contrast, clump size, and  time of initiation of the interaction
are the major factors. 
Ram pressure stripping gives rise to a core-plume structure 
and a strong
Kelvin-Helmholtz instability in the downstream region. 
The Rayleigh-Taylor instability  develops on the clump's upstream side,
which facilitates  fragmentation of the clump. The clump
causes a bulge on the remnant outline as the  ram pressure
pushes material ahead.
The bulge eventually disappears as the clump is fragmented and 
swept back. 
Punching through the forward shock
does not occur for our similution with conditions like those
in Tycho's remnant, even for the case of a clump with an initial
density contrast of 100.
Clump interaction in the early phases of the remnant evolution are more
likely to distort the forward shock front because of the larger density
contrast between the ejecta and the interstellar medium.

The two X-ray clumps to the SE in Tycho are probably not alone.
In fact, we find that the large radial position of most of the
ejecta emission suggests that much of the emission is from
clumps.
If it were not in clumps, it could not extend to the observed radius.
We thus suggest that the undecelerated knots are at the high column
density end of a spectrum of knot properties.
The required knot compression factor, up to $\sim 100$, is not
a natural result of existing Type Ia supernova models.
We suggest that the expansion of a radioactive Nickel bubble is
responsible for the compression of the knots.
The compression factor is larger than that found in existing
calculations, but it is possible that radiative transfer effects
during the final stages of the Ni bubble expansion can lead to
enhanced compression.
The Ni bubble interpretation has the implication that Fe clumps,
which are observed, cannot have been synthesized as $^{56}$Ni.
They must have a non-radiogenic origin, such as $^{54}$Fe.
The synthesis of high velocity matter in which $^{54}$Fe is the
dominant component has yet to be shown in computations of
nucleosynthesis.

In addition to these suggested theoretical investigations, advances
can be expected on the observational front.
Spatially resolved X-ray spectra of Type Ia supernova remnants, as
is possible with the {\it Chandra} and {\it XMM} observatories,
will be valuable for showing whether our picture of ejecta
clumps is valid.
We predict that the outer ejecta emission in Tycho's remnant has
a clumpier structure than the inner emission observed in the
Fe K line.


C.-Y. Wang thanks John Blondin and John Hawley for their
assistance with the numerical simulations, and Fred Lo and Academia Sinica
for providing their facilities.
R.A.C. is grateful to Vikram Dwarkadas, Peter H\"oflich, Phil Pinto,
and Craig Wheeler for useful discussions and correspondence.
The computations were carried
out on the  Cray T90 of NPACI, and the IBM SP2 
at University of Virginia. 
Support for this work was provided in part by NASA grant NAG5-8232.

\clearpage


\clearpage
\figcaption{evo.gif}{Series of density images illustrating the time
evolution of the dynamical instability.
The perturbation was imposed at $t'=0.00054$ with 1\% amplitude and an
$l=100$ mode of the spherical harmonic function.
The grid has 500 radial zones by 2000 angular zones in 1/2 of
a quadrant.
The plots at early stages display the central 1000 angular zones in
1/4 of a quadrant,
while the plots shown for later stages display 1/2 of a quadrant.
The contour levels are exponentially spaced between the
lowest value and the highest one sampled in the grid domain.}
\label{fig.evo}

\clearpage

\clearpage

\clearpage
 
\clearpage
\figcaption{resol.gif}
{Series of density contours with varying resolutions for four epochs:
$t'=0.0016$, $0.0027$, $0.011$, and $0.11$.
The lowest resolution uses 200 angular zones on 1/2
of a quadrant and the highest uses 2000 zones on 1/2 of a quadrant.
The top two low resolution plots exhibit the same dominant modes
on 1/2 of a quadrant, growing from the initial
perturbation modes $l=100$ and $l=50$.
The other plots are shown on 1/4 of a quadrant.
All of the plots use a 10\% perturbing amplitude.}
\label{fig.resol}

\figcaption{G0mplt.gif}
{Snapshots of a clump in the ejecta expanding into the shocked region.
The clump was initiated at $t'=0.207$,
with an initial density contrast $\chi=100$ and a size $a_0=1/8$.
The clump becomes crescent shaped and
causes a protrusion in the forward shock front.
The grid has 500 radial by 500 angular zones on 1/2 of a quadrant.}
\label{fig.G0mplt}

\figcaption{k2mplt.gif}{Snapshots of a clump with an initial density contrast
$\chi=100$ and a size $a_0=1/6$ in the ejecta expanding into the shocked region.
The interaction was initiated at  $t'=0.866$.
The Rayleigh-Taylor instability can be seen on the clump's downstream side.
The grid has 500 radial by 500 angular zones on 1/2 of a quadrant.}
\label{fig.k2mplt}

\figcaption{band.gif}
{Snapshots of multiple clumps with a initial density contrast
$\chi=50$ and a size $a_0=0.15$ in the ejecta expanding into the shocked
region.
The interaction was initiated at  $t'=0.866$.
The grid has 500 radial by 500 angular zones on 1/2 of a quadrant.
The top plot has one row of clumps and the bottom has two rows.}
\label{fig.band}

\clearpage

 \begin{table}[h]
 \begin{center}
 \caption{Clump Simulation Parameters \label{tabclump}}
 \begin{tabular}{lllll}
 \tableline
 \tableline
 $t'$ \tablenotemark{a} & $\rho_c$ \tablenotemark{b} &
 $R_{rs}$ \tablenotemark{c} & $R_{fs}$ \tablenotemark{d} &
 $dR$ \tablenotemark{e} \cr
 \tableline
 0.011  &  15.5 & 0.042 & 0.048 & 0.006  \\
 0.217  &  2.03 & 0.797 & 0.958 & 0.161 \\
 0.866  &  0.580  & 1.70 & 2.26 & 0.478  \\
 1.24  &  0.315  & 1.93 & 2.70 & 0.768  \\
 \tableline
 \end{tabular}
 \end{center}
 \tablenotetext{a}{starting time of the clump-reverse shock interaction}
 \tablenotetext{b}{clump density for $\chi=1$, normalized
 to the unshocked ISM}
 \tablenotetext{c}{normalized radius of the reverse shock }
 \tablenotetext{d}{normalized radius of the forward shock }
 \tablenotetext{e}{normalized intershock width}

 \end{table}


\clearpage



\begin{thebibliography} {}


\bibitem[Baade(1945)]{ba45}
Baade, W. 1945, \apj, 102, 309

\bibitem[]{}
Basko, M. 1994, ApJ, 425, 264

 
\bibitem[]{}
Branch, D., Doggett, J. B., Nomoto, K., 
 \& Thielemann, F.-K. 1985, \apj, 294, 619


\bibitem[Chevalier et al.(1981a)]{che82}
Chevalier, R. A.  1981a, \apj, 246, 267

\bibitem[Chevalier et al.(1981b)]{che82}
Chevalier, R. A.  1981b, Fund. of Cosmic Phys., 7, 1

\bibitem[Chevalier et al.(1982a)]{che82a}
Chevalier, R. A.  1982a, \apj, 258, 790

\bibitem[Chevalier et al.(1982b)]{che82b}
Chevalier, R. A.  1982b, \apj, 259, L85

\bibitem[Chevalier et al.(2000)]{che82}
Chevalier, R. A.  2000, in Astronomy with Radioactivities, ed. R. Diehl
(Garching: MPI), in press

\bibitem[Chevalier et al.(1992)]{cbe92}
Chevalier, R. A., Blondin, J. M., \& Emmering, R. 
T.  1992, \apj, 392, 118.

\bibitem[Dwarkadas]{d98}
Dwarkadas, V. V. 2000, \apj, submitted

\bibitem[Dwarkadas and Chevalier(1998)]{dc98}
Dwarkadas, V. V., \& Chevalier, R. A.  1998, \apj, 497, 807 (DC98)

\bibitem[]{}
Fesen, R. A., Wu, C.-C., Leventhal, M., \& Hamilton, A. J. S.
1988, ApJ, 327, 164
 
\bibitem[Fryxell et al.(1991)]{fma91}
Fryxell, B., M\"uller, E., \& Arnett, W. D. 1991, \apj, 367, 619


\bibitem[]{}
Hamilton, A. J. S., Sarazin, C. L., \& Szymkowiak, A. E. 1986, ApJ, 300, 713


\bibitem[]{}
Hamilton, A. J. S., Fesen, R. A., Wu, C.-C., Crenshaw, D. M., 
\& Sarazin, C. L. 1997, ApJ, 482, 838

\bibitem[]{}
Harkness, R. P. 1991, in SN 1987A and Other Supernovae, ed. I. J. Danziger
\& K. Kj\"ar (Garching: ESO), 447

\bibitem[]{}
H\"oflich, P. 2000, priv. commun.

\bibitem[]{}
H\"oflich, P., \& Khokhlov, A. 1996, ApJ, 457, 500

\bibitem[]{}
H\"oflich, P., Khokhlov, A. M., \& Wheeler, J. C. 1995,
ApJ, 444, 831

\bibitem[]{}
Hughes, J. P. 1997, in X-Ray Imaging and Spectroscopy of Cosmic Hot
Plasmas, ed. F. Makino and K. Mitsuda (Tokyo: Universal Academy Press), 359

\bibitem[]{}
Hwang, U., \& Gotthelf, E. V. 1997, ApJ, 475, 665

\bibitem[]{}
Hwang, U., Hughes, J. P., \& Petre, R. 1998, ApJ, 497, 833


\bibitem[]{}
Iwamoto, K., Brachwitz, F., Nomoto, K., Kishimoto, N., Umeda, H., Hix, 
W. P., \& Thielemann, R.-K. 1999, \apjs, 125, 439

\bibitem[]{}
Jun, B.-I., \& Jones, T. W. 1999, \apj, 511, 774

\bibitem[]{}
Jun, B.-I., \& Norman, M. L. 1996, \apj, 465, 800

\bibitem[]{}
Jun, B.-I., \& Norman, M. L. 1996b, \apj, 472, 245

\bibitem[]{}
Kane, J., Drake, R. P., \& Remington, B. A. 1999, \apj, 511, 335

\bibitem[]{}
Kane, J., Arnett, D., Remington, B. A., Glendinning, S. G.,
Baz\'an, G., M\"uller, E., Fryxell, B. A., \& Teyssier, R. 2000, \apj, 528, 989

\bibitem[Kamper(1978)]{kam78}
Kamper, K. W., \& van den Bergh, S. 1978, \apj, 224, 851


\bibitem[]{}
Khokhlov, A. 1991, \aap, 245, L25


\bibitem[]{}
Klein, R. I., McKee, C. F., \& Colella, P. 1994, \apj, 420, 213

\bibitem[]{}
Li, H., McCray, R., \& Sunyaev, R. A. 1993, ApJ, 419, 824

\bibitem[]{}
Livne, E., \& Arnett, D. 1995, 452, 62


\bibitem[]{}
Mazzali, P. A., Chugai, N., Turatto, M., Lucy, L. B., Danziger, I. J.,
Cappellaro, E., Della Valle M., \& Benetti, S. 1997, MNRAS, 284, 151

bibitem[Fryxell et al.(1991)]{fma91}
M\"uller, E., Fryxell, B., \& Arnett, D. 1991, \aap, 251, 505

\bibitem[]{}
Nomoto, K., Thielemann, F.-K., \& Yokoi, K. 1984, ApJ, 286, 644

\bibitem[Raymond(1984)]{ram84}
Raymond, J. C. 1984, ARA\&A, 22, 75

\bibitem[]{}
Reynoso, E. M., Moffett, D. A., Goss, W. M., Dubner, G. M., Dickel, J. R.,
Reynolds, S. P., \& Giacani, E. B. 1997, ApJ, 491, 816

\bibitem[Reynoso et al.(1997)]{rey97}
Reynoso, E. M., Vel\'azquez, P. F., Dubner, G. M., \& Goss, W. M.
1999, AJ, 117, 1827

\bibitem[]{}
Schneider, M., Dimonte, G., \& Remington, B. A. 1998, Phys. Rev. Lett., 80, 3507
 
\bibitem[]{}
Schaefer, B. E. 1996, ApJ, 459, 438

\bibitem[]{}
Seward, F., Gorenstein, P., \& Tucker, W. 1983, ApJ, 266, 287

\bibitem[Stone and Norman (1992)]{st92}
Stone, J., \& Norman, M. 1992, ApJS, 80, 753.

\bibitem[Strom et al.(1982)]{str82}
Strom, R. G., Goss, W. M., \& Shaver, P. A. 1982, MNRAS, 200, 473.

\bibitem[Tan et al.(1985)]{tan85}
Tan, S. M., \& Gull, S. G. 1985, MNRAS, 216, 949

\bibitem[]{}
Thielemann, F.-K., Nomoto, K., \& Yokoi, K. 1986, \aap, 158, 17

\bibitem[]{}
Truelove, K., \& McKee, C. F. 1999, \apjs, 120, 299

\bibitem[]{}
van den Bergh, S. 1993, ApJ, 413, 67

\bibitem[Vancura et al.]{van95}
Vancura, O., Gorenstein, P., \& Hughes, J. P. 1995, \apj, 441, 680

\bibitem[Velazquez et al.(1998)]{vel98}
Vel\'azquez, P. F., G\'omez, D. O., Dubner, G. M., Gim\'enez de Castro, G.,
\& Costa, A., 1998, \aap , 334, 1060

\bibitem[]{}
Whelan, J., \& Iben, I., Jr. 1973, ApJ, 186, 1007

\bibitem[]{}
Woosley, S. E. 1988, ApJ, 330, 218

\bibitem[]{}
Woosley, S. E., \& Weaver, T. A. 1994, ApJ, 423, 371

\bibitem[]{}
Wu, C.-C.,  Leventhal, M.,
 Sarazin, C. L. \& Gull, T. R. 1983, ApJ, 269, L5

\bibitem[]{}
Wu, C.-C., Crenshaw, D. M., Hamilton, A. J. S., Fesen, R. A., Leventhal, M.,
\& Sarazin, C. L. 1997, ApJ, 477, L53

\bibitem[]{}
Xu, J., \& Stone, J. M. 1995, \apj, 454, 172

\end{thebibliography}
\end{document}